\documentclass[twocolumn,showpacs,preprintnumbers,amsmath,amssymb,footinbib]{revtex4-1}
\usepackage[utf8]{inputenc}
\usepackage[english]{babel}
\usepackage{lmodern}
\usepackage{amsmath}
\usepackage{graphicx}
\usepackage{amssymb}
\usepackage{epstopdf}
\usepackage{ upgreek }
\usepackage{color}
\usepackage{setspace}
\usepackage{bbm}
\newcommand\varpm{\mathbin{\vcenter{\hbox{%
  \oalign{\hfil$\scriptstyle+$\hfil\cr
          \noalign{\kern-.3ex}
          $\scriptscriptstyle({-})$\cr}%
}}}}
\graphicspath{{Figures/}}
\usepackage[toc,page]{appendix}
\usepackage{tocloft}
\def\ii{{\rm i}}  
\begin{document}

\title[]{Coherent control in atomic chains: to trap and release a traveling excitation}

\author{R.~Guti\'{e}rrez-J\'{a}uregui}
\email[Email:]{r.gutierrez.jauregui@gmail.com}
\author{A.~Asenjo-Garcia}
\email[Email:]{ana.asenjo@columbia.edu}
\affiliation{Department of Physics, Columbia University, New York, NY, USA.}

\date{\today}
\begin{abstract}
We introduce a protocol for dynamical dispersion engineering in an atomic chain consisting of an ordered array of multi-level atoms with subwavelength lattice constant. This chain supports dark states that are protected from dissipation in the form of photon emission and can be understood as propagating spin waves traveling along the array. By using an external control field with a spatially-varying elliptical polarization we correlate internal and external degrees of freedom of the array in a controllable way. The coherent control over the atomic states translates into control over the group velocity of the spin waves. A traveling excitation can be stored and released without dissipation by adiabatically changing the control field amplitude. This protocol is an alternative to the more conventional electromagnetically-induced transparency, and exemplifies the rich physics born of the interplay between coherent control and correlated decay.
\end{abstract}
\maketitle

Interfacing light with matter endows photons with matter-like properties, such as a finite effective mass or a slow group velocity. A striking example is that of electromagnetically induced transparency (EIT)~\cite{Harris_1991a,Harris_1991b,Harris_1992,Hau_1999}, where laser pulses traveling inside a disordered atomic gas can be slowed down significantly. In EIT, a transparency window is opened into an otherwise optically dense medium by means of an external control field that generates a dark state due to destructive interference. Light propagates through the medium as a polariton~\cite{Juzeliunas_1996} whose speed is reduced as the excitation is predominantly transferred to the dark state of the atomic gas. Moreover, this polariton can be brought to a stand-still by using a time-varying control field~\cite{Juzeliunas_2002,Fleischhauer_2002}. These ideas have been exploited for storing and releasing single photon states and form the backbone of quantum memories,  a key ingredient for information processing and communication.

Ordered arrays of atoms with subwavelength inter-atomic distances have emerged as an efficient light-matter interface, where interference between radiative paths leads to a collective response and photon emission can be enhanced or suppresed~\cite{Bettles_2016,Shahmoon_2017, Yoo_2020, Bloch_2020,Utyushev_2021}. Interest in this field has recently moved past the theoretical realm into experimental reality, with the realization of a two-dimensional atomic mirror~\cite{Bloch_2020}. Correlated decay has implications for the performance of quantum applications (such as quantum memories for light~\cite{Asenjo_2017}, photonic quantum gates~\cite{cardoner}, or atomic clocks~\cite{Kramer_2016,Henriet_2019}), which should no longer be limited by single-atom spontaneous emission.

In one-dimensional (1D) arrays, single-excitation dark states can be understood as guided modes of an atomic waveguide. They provide a natural transparency window that allows for lossless transport~\cite{Masson_2020}. Transparency does not arise from destructive interference between different internal states of a single atom (like in EIT), but from collective interference in the radiated field. Dissipation in the form of photon emission is thus suppressed, with subradiant excitations being bound to the array~\cite{Asenjo_2017}.

Here, we present a protocol for dynamical dispersion engineering in 1D arrays that allows for trapping and releasing single traveling excitations.  We use an external control field that is far-detuned from the excited states of atoms with a three-level $V$ configuration. The field displays a subwavelength polarization gradient that modifies the dipole-dipole interactions between the atoms and effectively couples their internal (i.e., spin) and external (i.e., position) degrees of freedom. Coherent control over the internal atomic states is translated into control over the group velocity of an excitation that propagates through the array. Adiabatic changes of the helicity of the dressing fields results in an effective ``Paul trap'' for photons, allowing to slow down, trap, and reverse the direction of traveling excitations, all without dissipation.

\begin{figure}[h]
\begin{center}
\includegraphics[width=.85\linewidth]{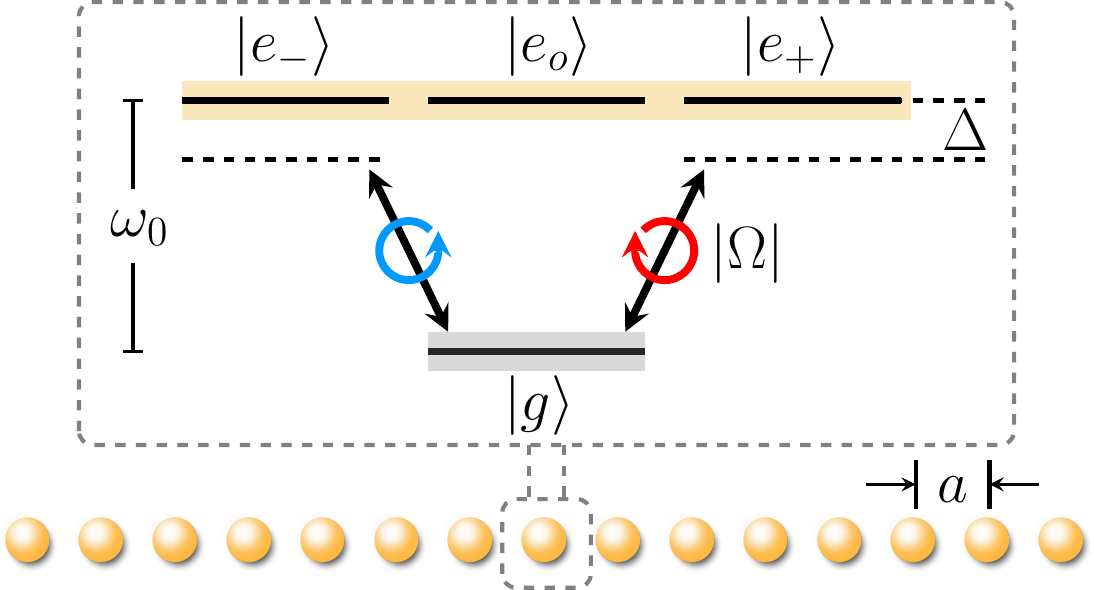}
\includegraphics[width=.85\linewidth]{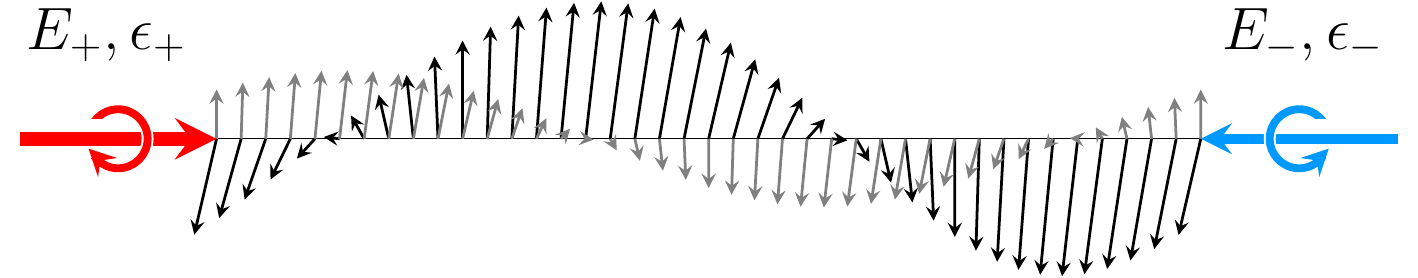}
\caption{An atomic chain of lattice constant $a$ is driven by a far-detuned control field with periodic polarization gradient. The field consists of two counter-propagating circularly-polarized plane waves that form a standing wave of elliptical polarization whose axes rotate along the chain axis (black and gray arrows) and induces a position-dependent dipole moment by coupling ground $\vert g \rangle$ and excited $\vert e _{\pm} \rangle$ states. The coupling strength depends on the Rabi frequency $\vert \Omega \vert$ and detuning $\Delta$ with a relative phase that depends on the atomic position. The field enables control over the dispersion relation of dark states that emerge in the chain for lattice constants $a<\lambda_0/2$ ($\lambda_0$ being the atomic transition wavelength).} \label{Fig:energy_levels} 
\end{center}
\end{figure}

The proposed set-up, sketched in Figure~\ref{Fig:energy_levels}, consists of a chain of lattice constant $a$ where each atom is characterized by its position $z_{n}$ and is assumed to have a ground state $\vert g^{n} \rangle$ and three excited states $\vert e_{s}^{n}\rangle$ ($s= \lbrace 0,\pm \rbrace$). The chain is illuminated by a control field composed of two counter-propagating plane-waves of electric field amplitudes $E_{+}$ and $E_{-}$ that share the same frequency $\omega_{c}$, but have counter-rotating circular polarizations $\mathbf{e}_{\pm}$. Their superposition creates a standing wave of elliptical polarization and constant ellipticity throughout the chain~\cite{Dalibard_1989},
\begin{equation}
\mathbf{E}_{c}(\mathbf{z},t) = e^{-\ii \omega_{c}t} \lbrace (E_{+} + E_{-}) \mathbf{e}_{x}^{\prime} + \ii (E_{+} - E_{-}) \mathbf{e}_{y}^{\prime} \rbrace + c.c. \, , \nonumber
\end{equation}
whose axes $\mathbf{e}_{x,y}^{\prime}$ rotate around the chain axis
\begin{align}
\mathbf{e}_{x}^{\prime} = \cos k_{c} z \,\mathbf{e}_{x} - \sin k_{c} z \,\mathbf{e}_{y} \, ,\nonumber \\
\mathbf{e}_{y}^{\prime} = \sin k_{c} z \,\mathbf{e}_{x} + \cos k_{c} z \,\mathbf{e}_{y} \, .\nonumber
\end{align}  

We first discuss how the control field induces a position-dependent dipole moment in the atoms due to its polarization gradient. Each atom couples to this field via the dipolar interaction $\hat{\mathbf{d}}^{(n)}\cdot \mathbf{E}_{\text{c}}(z_{n})$ where $\hat{\mathbf{d}}^{(n)}$ is the electric dipole moment operator and $\mathbf{E}_{\text{c}}(z_{n})$ the field amplitude evaluated at the atomic position $z_n$. The coupling strength $\hbar\Omega = \sum_{s} d_{s}E_{s}(z_{n})$ is determined by the electric-dipole matrix elements between states $g$ and $e_{s}$, denoted by $d_{s}$, and the projection of the field along their respective orientations. With the control field far-detuned from the atomic transition frequency $\omega_{0}$ ($\Delta \gg \vert \Omega \vert$, with $\Delta=\omega_0-\omega_c$), its effect on the $n$th-atom is accurately described by the Hamiltonian~\cite{Messiah_1961,Mitroy_2010,supp_us}
\begin{align}
\hat{\mathcal{H}}^{(n)}_{\text{\scriptsize{eff}}} =& \frac{\hbar}{2}(\Delta+\delta)(\sum_{s=\pm}\hat{\sigma}^{(n)}_{ss}-\hat{\sigma}^{(n)}_{gg} )- \sum_{s = \pm} \frac{\hbar\delta}{4}(1-s \cos \theta)\hat{\sigma}_{ss}^{(n)} \nonumber \\
+& \frac{\hbar \delta}{4} \sin \theta \left(e^{-2ik_{c}z_{n}}\hat{\sigma}^{(n)}_{\scriptsize{+ -}} + e^{2ik_{c}z_{n}}\hat{\sigma}^{(n)}_{\scriptsize{- +}} \right), 
\end{align}
plus an additional term accounting for the state $\vert e^{n}_{o} \rangle$, which remains decoupled from the control field and is ignored in what follows. In this expression, the transfer operator $\hat{\sigma}_{s s^{\prime}}^{(n)} = \vert e_{s}^{n}\rangle \langle e_{s^{\prime}}^{n}\vert$ connects two atomic states, and the parameter $\delta = \vert \Omega \vert^{2}/2\Delta$ represents the frequency shift induced by the light. The local polarization of the field is imprinted on each atom through the mixing angle
\begin{equation}
\theta = 2 \arctan (E_{+}/E_{-}) \, ,
\end{equation}
and the phase $\pm 2 k_{c}z_{n}$. The phase carries all the information regarding the atomic position. For atomic ensembles confined to regions much smaller than their transition wavelength this phase is irrelevant and can be absorbed into a global coupling parameter. It, however, plays a central role on extended systems~\cite{Lehmberg_1970}, as we now discuss for an atomic chain.

Atoms in an ordered chain interact with one another via the modes of the surrounding electromagnetic environment. Tracing out these modes under the Born and Markov approximations gives rise to a master equation for a density matrix that describes atomic degrees of freedom only~\cite{Gross_1982, Carmichael_1999}. As we are interested in single-excitation dynamics we consider a quantum trajectory evolution equivalent to the resulting master equation. The atomic state is then described by an ensemble of stochastic wavefunctions evolving under a Schr\"{o}dinger equation with non-Hermitian Hamiltonian 
\begin{equation}\label{eq:non_hermitian}
\tilde{\mathcal{H}} = \sum_{n}\hat{\mathcal{H}}^{(n)}_{\text{eff}} - \hbar \sum_{n,m=1}^N \sum_{s=\pm} \mathcal{K}_{s,s}^{n,m} \hat{\sigma}_{sg}^{(n)} \hat{\sigma}_{gs}^{(m)} \, ,
\end{equation} 
interrupted by jump operators that drive the system into the absolute ground state when an excitation leaves the array. In the above equation, the coupling constants read~\cite{Gross_1982, Carmichael_1999,Asenjo_2017}
\begin{equation}\label{eq:nueva_Green}
\mathcal{K}_{s, s}^{n,m} = \frac{3\pi\Gamma_{0}}{k_0}  {\mathbf{e}}^{*}_{s}\cdot \mathbf{G}(z_{n}-z_{m},\omega_0) \cdot {\mathbf{e}}_{s} \, ,
\end{equation}
where $\mathbf{G}(z_{n}-z_{m},\omega_0)$ is the propagator of the electric field scattered from site $n$ to site $m$ and $\mathbf{e}_{s}$ the atomic transition polarization of the involved states~\cite{green_us}; $\Gamma_{0}$ refers to the single atom spontaneous emission rate, and $k_0=\omega_0/c$ is the wave-vector associated to the atomic transition frequency. Without the elliptical control field, the Hamiltonian is diagonal in polarization indices as the field propagator does not mix different polarization components along the chain direction~\cite{Asenjo_2019}. 

\begin{figure*}
\begin{center}
\includegraphics[width=.95\linewidth]{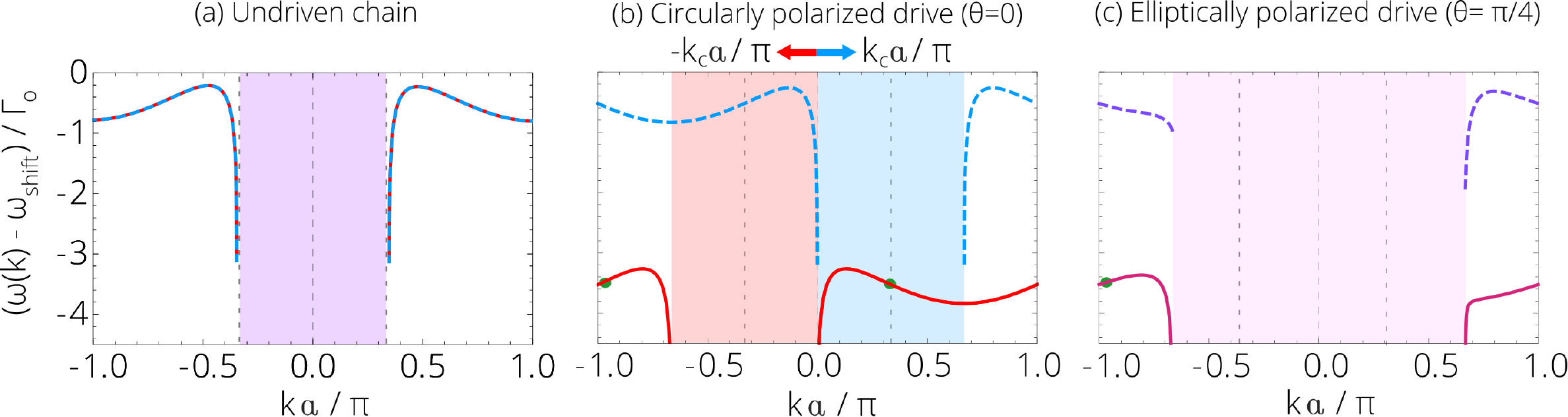}
\caption{Dispersion engineering by an external control field. (a) Dispersion relation of an undriven chain, where the two bands corresponding to the two internal degrees of freedom ($\sigma_\pm$-transitions) are degenerate. (b, c) Dispersion relation of upper (dashed line) and lower (solid line) states of a chain driven by a (b) circularly-polarized and (c) elliptically-polarized control field. Red and blue color represents the relative weight of the $|e_{+}\rangle$ and $|e_{-}\rangle$ transitions in the polariton, respectively. Notice both states mix for elliptically an polarized drive. For all plots the lattice constant is  $a=\lambda_0/6$, and the radiative region is shaded. In (a) the frequency shift is $\omega_\text{shift}=\omega_0$. In (b, c), $\omega_{\text{shift}} =\omega_0+ (2\Delta + \delta)/4$, the control field has wavelength $\lambda_{c} = \lambda_0$ and generates a light-shift $\delta = 6 \Gamma_0$. Green dots indicate the central frequency of the excitation propagating in Figs.~\ref{Fig:pulse_1} and~\ref{Fig:dispersion}.} \label{Fig:energy_bands}
\end{center}
\end{figure*}

The control field introduces a local rotation of the polarization that mixes different excited states along the chain. The system is invariant under a displacement of a lattice constant $a$ together with a rotation of an angle $k_c a$. This helical symmetry is represented by the operator
\begin{equation}
\hat{I}(a,k_{c}) = \exp \left[\frac{\ii \hat{P}_{z}a + \ii \hat{J}_{z} k_{c} a}{\hbar}\right] \,,
\end{equation}
which tracks the local polarization of the control field along the array, with the linear momentum operator $\hat{P}_{z}=-i\hbar \partial_z$ generating displacements along the chain  and the angular momentum operator $\hat{J}_{z}= \hbar [\sum_n \hat{\sigma}_{++}^{(n)} - \hat{\sigma}_{--}^{(n)}]$ generating rotations around the chain axis. This operator commutes with the Hamiltonian and provides a basis to diagonalize it. The eigenstates of the Hamiltonian are then found to be Bloch waves of quasimomentum $k \in [-\pi/a,\pi/a)$~\cite{supp_us}:

\begin{widetext} 
\begin{subequations}\label{eq:eigenstates}
\begin{align}
\vert {\text{{U}}}, k \rangle & = \sum_{n} e^{\ii k z_{n}} \left[ e^{\ii k_{c}z_{n}} \sin (\alpha_k/2) \,\hat{\sigma}_{-g}^{(n)} + e^{-ik_{c}z_{n}} \cos  (\alpha_k/2) \,\hat{\sigma}_{+g}^{(n)} \right]  \vert g \rangle^{\otimes N} \, , \\
\vert {\text{{L}}}, k \rangle & = \sum_{n} e^{\ii k z_{n}} \left[e^{\ii k_{c}z_{n}} \cos  (\alpha_k/2)\, \hat{\sigma}_{-g}^{(n)} - e^{-ik_{c}z_{n}} \sin  (\alpha_k/2) \,\hat{\sigma}_{+g}^{(n)} \right] \vert g \rangle^{\otimes N} \, ,
\end{align}
\end{subequations}
where the index $\{\text{U},\text{L}\}$ accounts for the internal degrees of freedom~\cite{Castin_1991,Marte_1993,Ren_1995} and $\alpha_k$ represents the relative populations of excited states~\cite{us_2}. The dispersion relation $\omega_{\text{\tiny{U,L}}}$ and collective decay $\Gamma_{\text{\tiny{U,L}}}$ of the ``upper'' and ``lower'' states   satisfy
\begin{equation}\label{eq:eigenvalues}
\omega_{\text{\tiny{U(L)}}}(k) - \tfrac{\ii}{2}\Gamma_{\text{\tiny{U(L)}}}(k) =  \tfrac{1}{2}\left[\Delta + \tfrac{1}{2}\delta -\ii \Gamma_{0} - \tilde{\mathcal{K}}(k+k_{c}) - \tilde{\mathcal{K}}(k-k_{c}) \varpm  \tfrac12 \Omega_k\right]. 
\end{equation}
\end{widetext}
In the above equation, $\tilde{\mathcal{K}}$ is the Fourier transform of Eq.~\eqref{eq:nueva_Green} for a specific polarization~\cite{Asenjo_2017} , i.e.,   $$\tilde{\mathcal{K}}(k) = \frac{3 \Gamma_{0}}{4\ii}  \sum_{m =1}^{3} \left(\frac{\ii}{a k_0}\right)^{m}[\text{Li}_{m}(e^{i(k_0+k)a}) + \text{Li}_{m}(e^{i(k_0-k)a})] $$
with $\text{Li}_{m}$ the polylogarithm of order $m$~\cite{Abramowitz} and $\Omega_k$ is the band splitting~\cite{supp_us}, which reads
\begin{equation}\nonumber
\Omega_k = \sqrt{(\delta \sin \theta)^{2} +  (\delta \cos \theta + 2 [\tilde{\mathcal{K}}(k+k_{c}) - \tilde{\mathcal{K}}(k-k_{c})])^{2}}.
\end{equation}

For $a<\lambda_0/2$, subradiant states with $\Gamma_\text{U,L}(k)=0$ emerge. These states can be identified as guided modes of the atomic waveguide and their dispersion relation is shown in Fig.~\ref{Fig:energy_bands}. Subradiant states lie beyond the light line (i.e., they have a wavevector such that $|k|>k_0$) and cannot decay radiatively due to energy-momentum mismatch. This is readily seen in Fig.~\ref{Fig:energy_bands}(a) where, in the absence of the control field, the guided modes associated to each excited state $|e_{\pm}\rangle$ are degenerate. In the language of electromagnetically induced transparency, one can understand guided modes as providing a transparency window for photons to travel along the array without being scattered out. Due to the near-field dipole-dipole interaction, the transparency bandwidth scales as $\sim\Gamma_{0}/(k_0 a)^{3}$. 

\begin{figure}
\begin{center}
\includegraphics[width=1.\linewidth]{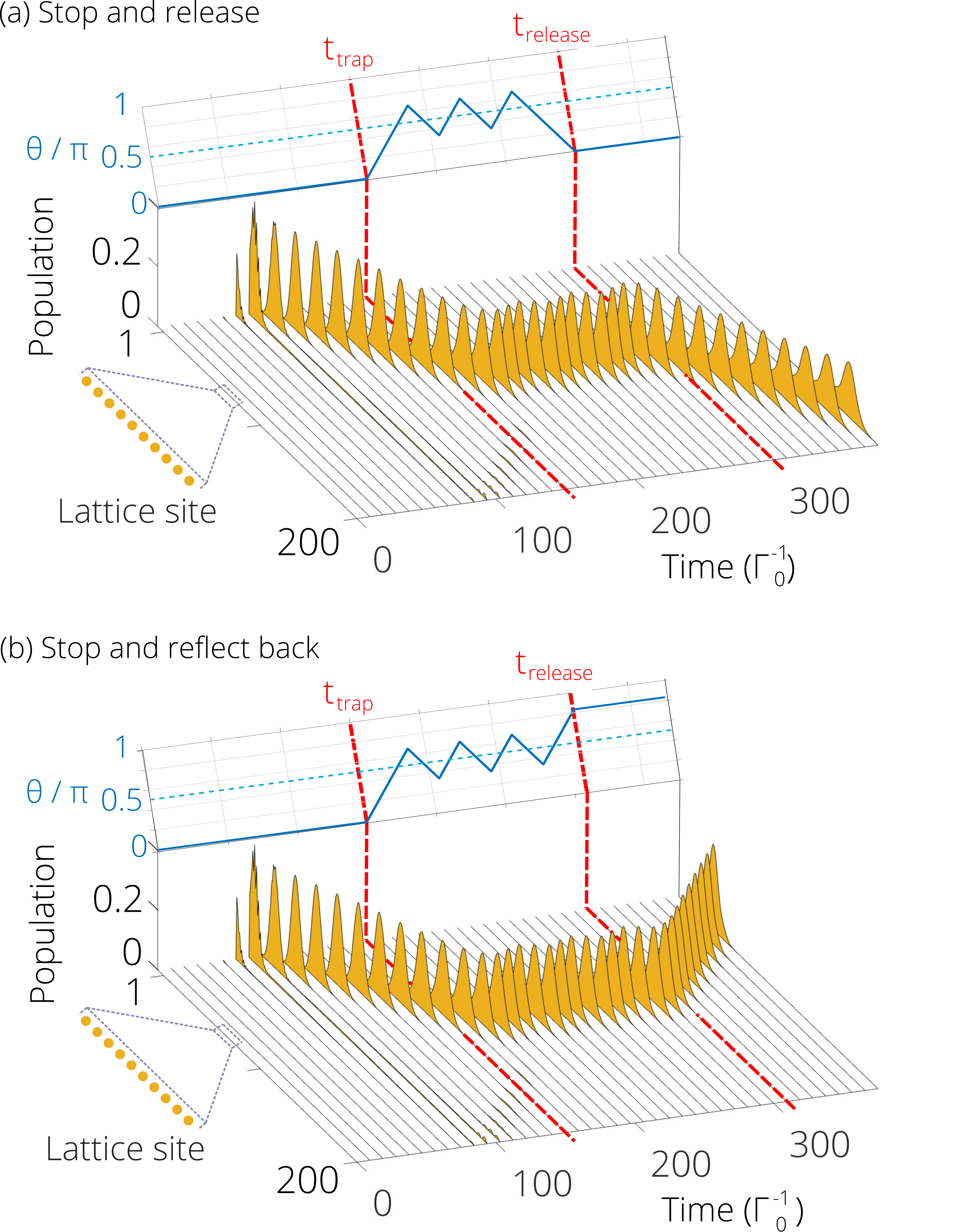}
\caption{Dynamical trapping of an excitation inside an atomic chain of $N=200$ atoms before it is released (a) or reflected (b). In both cases the excited state population of each site~(yellow distribution) and the mixing angle (blue line) are shown as a function of time. The initial chain parameters follow from Fig.~\ref{Fig:energy_bands}(b) while the driving field is described in the main text. } \label{Fig:pulse_1}
\end{center}
\end{figure}

The control field breaks the degeneracy between the excited states and introduces a relative phase between $|U,k\rangle$ and $|L,k\rangle$, which displaces the center of the light cone for each band by $\pm k_{c}$, as shown in Figs.~\ref{Fig:energy_bands}(b,c). For circular polarization [$\theta=0$, Fig.~\ref{Fig:energy_bands}(b)] there is no mixing between $|e_{\pm}\rangle$ states. The dispersion relation is identical to that of an undriven chain, except for a frequency shift and a displacement in quasimomentum (a shift from the origin). For elliptical polarizations [$\theta\neq0$, Fig.~\ref{Fig:energy_bands}(c)], there is coupling between internal (polarization) and external (position) degrees of freedom, resulting in the mixing of $|e^{n}_{+}\rangle$ and $|e^{m}_{-}\rangle$ [see Eqs.~(\ref{eq:eigenstates})]. 

The angle $\theta$ is controlled by the ratio between control-field amplitudes $E_\pm$ and determines the mixing of the bands. It also determines the group velocity of the excitations that propagate in the form of spin waves. The control field governs the dispersion relation of the chain, and allows us to, for instance, modify the bandwidth (by approaching $\theta=\pi/2$, where coupling between neighboring sites is reduced and the bands flatten), or to make the system non-reciprocal~\cite{Hadad_2010}.

The group velocity can be changed mid-flight by using a time-varying control field. Consider an excitation with a central frequency that lies in the lower band $\omega_{\text{L}}$ and has an initial group velocity $v_{\text{\tiny{L}}} = \partial \omega_{\text{\tiny{L}}}/\partial k $. Adiabatic changes~\cite{Dahan_1996,Peik_1997,Dalibard_2011} of the mixing angle allow for the coherent and reversible transfer of the excitation from states $|e_+\rangle$ to $|e_-\rangle$, while keeping the central quasi-momentum unperturbed. This transfer induces changes on the group velocity 
\begin{equation}
\dot{v}_{\text{\tiny{L}}} \propto  \dot{\theta} \sin \theta  \partial_{k} \left[\tilde{\mathcal{K}}(k+k_{c}) -  \tilde{\mathcal{K}}(k-k_{c}) \right].
\end{equation}

Coherent control of the group velocity can be used to trap and release a pulse as it propagates along the chain. As an example, we simulate numerically an excitation created by driving the first atom in the chain using a linearly polarized Gaussian pulse with temporal width of $25 \Gamma_{0}^{-1}$ and central frequency detuned from the atomic transition by $\Delta_{p}= -3.5\Gamma_{0}$ corresponding to the green dots in Fig.~\ref{Fig:energy_bands}(b). Figure~\ref{Fig:pulse_1} shows the spin wave propagation as a function of time and lattice site. The evolution is conditioned to no excitations leaving the array. The chain is initially prepared in the ground state and is driven by a circularly polarized control field $(\theta=0)$ to guarantee a large bandwidth and low dispersion. At time $t_0= 50 \Gamma_{0}^{-1}$ the excitation is created. It propagates freely before the control field is adiabatically changed until it becomes linearly polarized $(\theta = \pi/2)$ bringing the pulse near to a standstill. The mixing angle is then varied to oscillate around $\theta = \pi/2$ for three cycles before it is released [$\theta=0$, Fig.~\ref{Fig:pulse_1}(a)] or reflected back [$\theta = 0.8\pi$, Fig.~\ref{Fig:pulse_1}(b)]. 

\begin{figure}
\begin{center}
\includegraphics[width=1.\linewidth]{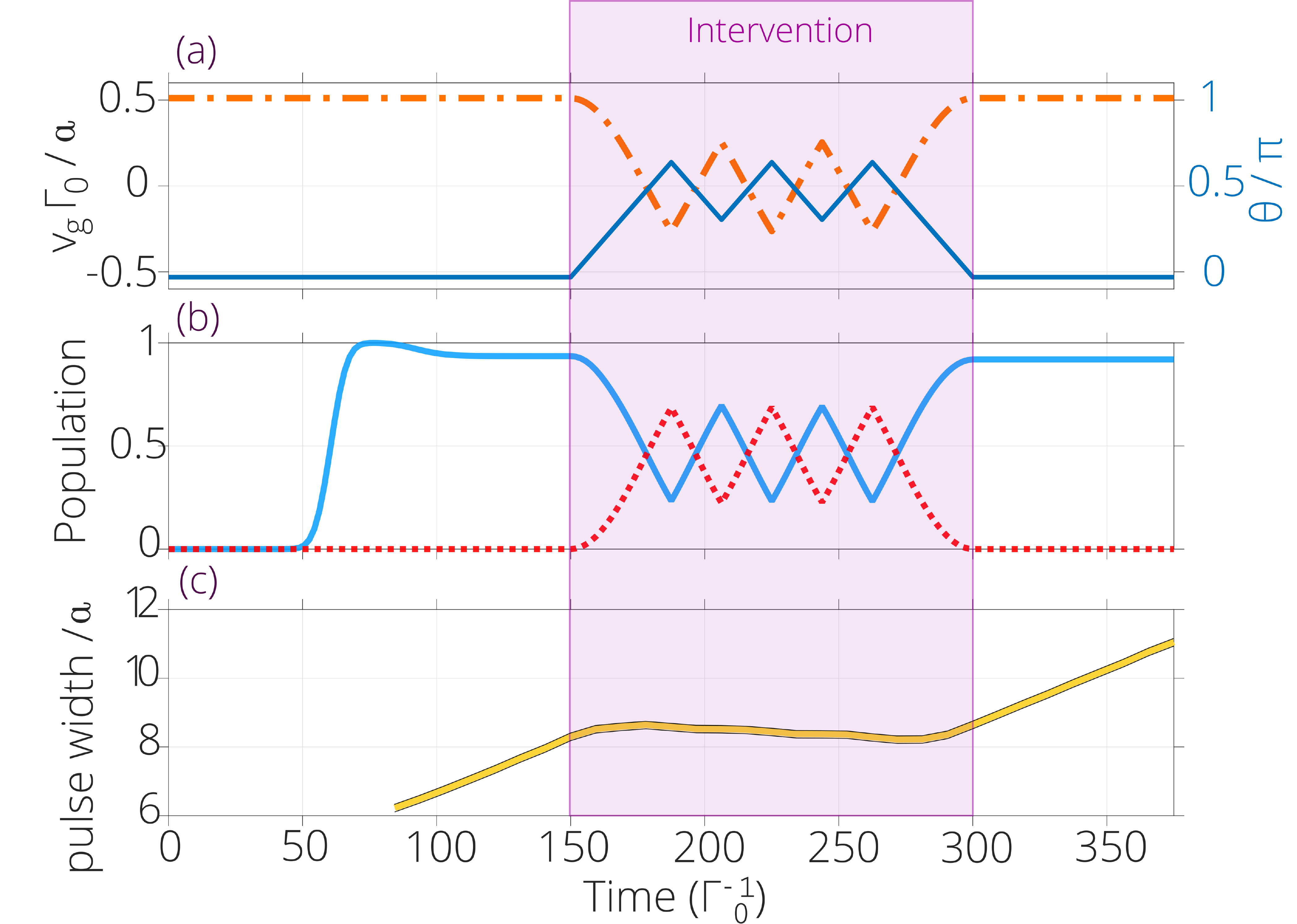}
\caption{Internal dynamics and dispersion control of the trapped excitation of Fig.~\ref{Fig:pulse_1}(a). (a) Group velocity (orange dashed line) and mixing angle (blue solid line). (b) Total population of the $\vert e_{-}\rangle$ (light blue solid) and $\vert e_{+}\rangle$ (red dotted) states. (c) The pulse width remains virtually unchanged during the intervention.} \label{Fig:dispersion}
\end{center}
\end{figure}

Dispersion limits the trapping time, but this effect can be corrected by the control field. The dispersion relations $\omega_{\text{L}}$ and $\omega_{\text{U}}$ display opposite curvatures in most of the subradiant region [see solid and dashed lines in Figs.~\ref{Fig:energy_bands}(b,c)]. Therefore, the phase acquired by each quasimomentum component changes from positive to negative as the ellipticity of the control field is varied. While this behavior can be exploited to trap the excitation, it can also be used to compensate for its natural dispersion. Figure~\ref{Fig:dispersion} shows the population of each atomic transition and the width of the pulse as a function of time. The population is normalized to the coupling efficiency (a maximum excited state population of~$\simeq 0.2$ is reached for our set of parameters). The width of the wave-packet is approximated by the mean square-root deviation obtained from a gaussian fit to the excitation probability. As the ellipticity of the control field changes, there is a partial rephasing of different quasimomentum components, thus reducing the dispersion. 

In contrast to EIT~\cite{Juzeliunas_2002,Fleischhauer_2002}, where the storage time is limited by the decay of the metastable state, our protocol is not constrained by the excited state lifetime as the decay is suppressed collectively via subradiance. As in EIT, the storage procedure relies in coherent transfer of an excitation between two states. In EIT, the dark polariton mostly consists of ground state coherences. Here, the coherences involve the excited states, which resembles ``stationary EIT''~\cite{Andre_2002}.

The experimental realization of these ideas require stringent (but achievable) conditions on the hyperfine structure of the atomic species and on the lattice constant. In particular, it requires the availability of a $F_{g}=0$ to $F_{e}=1$ transition, found in bosonic Strontium and Ytterbium, and deeply subwavelength interatomic distances, which have been recently achieved in optical lattice setups~\cite{Wang_2018,Anderson_2020,Tsui_2020}. It also requires for the control field frequency to be similar to that of the atomic resonance, imposing that $k_c\simeq k_0$. Small lattice constants give rise to increased Brillouin zones (and larger bandwidths), and the relative shift in quasimomentum of the upper and lower bands decreases. That makes the two bands more similar, and it is difficult to find regions where they display opposite curvature, which is required to trap the excitation. To move past this restriction one can choose a control field that couples the excited states to a third, highly excited, state. The third state acts as a bridge between $\vert e_{\pm} \rangle$ while allowing for $k_{c}$ and $k_0$ to be independent from each other.

Dynamical control and dispersion engineering make atomic arrays light-matter interfaces of unprecedented versatility. We have demonstrated that it is possible to manipulate the optical properties of ordered atomic media via external control fields, which is hard to achieve in conventional dielectric structures. These ideas can be exploited to trap and release single-excitation states, opening the door to the development of novel protocols for quantum information processing. Long trapping times may enable the realization of two-photon gates without the need for Rydberg interactions. Moreover, the control field acts as a magnetic field that induces a local vector shift in the atoms. This may be utilized to engineer other exotic properties, such as non-reciprocal transport (where backscattering is inhibited and transport becomes robust to disorder).

\textbf{Acknowledgments - } We thank D.~E.~Chang, J.~A.~Muniz, H.~Ochoa, A.~Gonzalez-Tudela, R.~J\'{a}uregui, and L.~A.~Orozco for insightful comments and discussions. R.G.-J. and A.~A.-G. acknowledge financial support by the National Science Foundation QII-TAQS (Award No. 1936359).


\begin{thebibliography}{1}
\bibitem{us_2} The angle $\alpha_k$ is defined by \begin{equation}
\tan \alpha_k = \frac{\delta \sin \theta}{\delta \cos \theta + 2 [\tilde{\mathcal{K}}(k+k_{c}) - \tilde{\mathcal{K}}(k-k_{c})]} \, . \nonumber
\end{equation}
\end{thebibliography}
\end{document}


\title[]{Supplemental Material for: Coherent control in atomic chains: to trap and release a traveling excitation}

\author{R.~Guti\'{e}rrez-J\'{a}uregui}
\email[Email:]{r.gutierrez.jauregui@gmail.com}
\author{A.~Asenjo-Garcia}
\email[Email:]{ana.asenjo@columbia.edu}
\affiliation{Department of Physics, Columbia University, New York, NY, USA.}

\date{\today}
\begin{abstract}
The effective Hamiltonian describing the evolution of each atom in the presence of the control field and the eigenstates of the atomic chain in the infinite limit are derived.
\end{abstract}
\maketitle

\section{Effective Hamiltonian}

In this Appendix we derive the coherent response of each atom to the control field. This corresponds to $\mathcal{H}_{\text{eff}}^{(n)}$ in Eq.~(1) of the main text. We consider a perturbative approach valid when the control field is far detuned from the atomic transition that is standard in the literature of atomic polarizabilities~\cite{Mitroy_2010, Deutsch_2010}. 

The evolution of the $n$th atom coupled to the control field under the dipole approximation is described by the Hamiltonian
\begin{equation}
\mathcal{H}^{(n)} = \tfrac{1}{2} \hbar \omega_{0} (\hat{\mathcal{P}}^{(n)}_{e}-\hat{\mathcal{P}}^{(n)}_{g}) + \hat{\mathbf{d}}^{(n)}\cdot \mathbf{E}_{\text{c}}(z_{n}) \, ,
\end{equation}
with projectors $\hat{\mathcal{P}}^{(n)}_{g} = \vert g^{n} \rangle \langle g^{n} \vert$ and $\hat{\mathcal{P}}^{(n)}_{e} = {1}^{(n)} - \hat{\mathcal{P}}^{(n)}_{g}$ that act upon ground- or excited-state subspaces and control field amplitude $\mathbf{E}_{\text{c}}(\mathbf{z},t)$ that reads
\begin{equation}
\mathbf{E}_{\text{c}}(\mathbf{z},t) = e^{-i\omega_{c}t} \frac{\vert E\vert }{2}  ( e^{i k_{c}z}\cos \frac{\theta}{2} \mathbf{e}_{+} + e^{-i k_{c}z}\sin \frac{\theta}{2} \mathbf{e}_{-})  + c.c. \,  \nonumber
\end{equation}
when written in the circular polarization basis $(\mathbf{e}_{\pm} = \mathbf{e}_{x} \pm i \mathbf{e}_{y} )$. We apply the rotating wave approximation and move to an interaction picture with respect to the control field frequency where the dynamics are simplified. In this picture the interaction Hamiltonian takes the form
\begin{align}
\tilde{\mathcal{H}}^{(n)} = \tfrac{1}{2} \hbar\Delta (\hat{\mathcal{P}}^{(n)}_{e}-\hat{\mathcal{P}}^{(n)}_{g}) + V^{(n)}
\end{align}
with coupling potential
\begin{align}
V^{(n)} = \frac{\hbar \vert \Omega \vert }{2} ( e^{-i k_{c}z_{n}} \cos \frac{\theta}{2} \hat{\sigma}^{(n)}_{+g} +  e^{i k_{c}z_{n}} \sin \frac{\theta}{2} \hat{\sigma}^{(n)}_{-g} ) + H.c. \, . \nonumber
\end{align}
To obtain this equation we have taken into account that the electric dipole matrix elements connecting ground and excited states are equal for all the excited states ($d_{\pm} = d$) and have defined
\begin{align}
&\Delta = \omega_{0} - \omega_{c} \, , \\
&\hbar \vert \Omega \vert  = 2 \vert E \vert d \, , 
\end{align}
in accordance to the main text.

For far detuned fields $(\Delta \gg \vert \Omega \vert)$ the coupling $V^{(n)}$ can be taken as a perturbation that shifts the atomic energy levels. These light shifts $\epsilon_{l}$ can  be obtained using perturbation theory for degenerate states~\cite{Messiah_1961}. The second order contributions of these light shifts are given by solutions to the eigenvalue equation
\begin{equation}\label{matrix_eq}
\hat{\mathcal{P}}_{e(g)}^{(n)} \left[ (\hbar\Delta)^{-1} V^{(n)} \hat{\mathcal{P}}_{g (e)}^{(n)} V^{(n)} \right] \vert \phi_{e(g)}^{(0)} \rangle = \epsilon_{e (g)} \vert \phi_{e(g)}^{(0)} \rangle \, ,
\end{equation}
while the first-order contributions due to the spherical symmetry of the bare atomic states. Solving Eq.~(\ref{matrix_eq}) gives $\epsilon_{e1} = 0$, $\epsilon_{e2} = \delta/2 $, and  $\epsilon_{g} = -\delta/2$ with $$\delta = \vert \Omega \vert^{2}/2\Delta.$$ Notice that the level shifts add up to zero. 

This perturbative response is equivalent to an effective Hamiltonian where the coupling $V^{(n)}$ is replaced by the matrices acting on ground and excited subspaces that are defined in Eq.~(\ref{matrix_eq}). The effective Hamiltonian then reads
\begin{align}
\mathcal{H}^{(n)}_{\text{\scriptsize{eff}}} =& \frac{\hbar}{2}(\Delta+\delta)(\hat{\mathcal{P}}^{(n)}_{e}-\hat{\mathcal{P}}^{(n)}_{g}) - \sum_{s = \pm} \frac{\hbar\delta}{4}(1-s \cos \theta)\hat{\sigma}_{ss}^{(n)} \nonumber \\
+& \frac{\hbar \delta}{4} \sin \theta \left(e^{-2ik_{c}z_{n}}\hat{\sigma}^{(n)}_{\scriptsize{+ -}} + e^{2ik_{c}z_{n}}\hat{\sigma}^{(n)}_{\scriptsize{- +}} \right), 
\end{align}
as given in Eq.~(1) of the main text.

\section{Eigenstates of the collective non-Hermitian Hamiltonian}

In this Appendix we obtain the eigenstates and eigenvalues of the non-Hermitian Hamiltonian
\begin{equation}\label{eq:non_hermitian}
\tilde{\mathcal{H}} = \sum_{n}{\mathcal{H}}^{(n)}_{\text{eff}} - \sum_{n,m=1}^N \sum_{s=\pm} \hbar \mathcal{K}_{s,s}^{n,m} \hat{\sigma}_{sg}^{(n)} \hat{\sigma}_{gs}^{(m)} \, ,
\end{equation} 
for an atomic chain where one excitation is distributed along the chain. The derivation follows closely the work by Ren and Carmichael~\cite{Ren_1995} where the eigenstates of an atom inside a standing-wave cavity are obtained from a set of conserved operators. For the atomic chain these operators are the total excitation and helical operators. We now address their individual effects. 

Since the Hamiltonian commutes the excitation number
\begin{equation}
[ \tilde{\mathcal{H}}, \sum_{n}\sum_{s=\pm} \hat{\sigma}_{ss}^{(n)} ] = 0
\end{equation}
we focus into the single excitation subspace where the eigenstates take the form
\begin{equation}
\vert \phi \rangle = \sum_{n} \sum_{s=\pm} c_{s}^{(n)} \hat{\sigma}_{sg}^{(n)} \vert g \rangle^{\otimes N}
\end{equation}
with $c_{s}^{(n)}$ probability amplitudes. The helical operator, in turn, is defined as
\begin{equation}
\hat{I}(a,k_{c}) = \exp \left[\frac{\ii \hat{P}_{z}a + \ii \hat{J}_{z} k_{c} a}{\hbar}\right] \, 
\end{equation}
and is responsible of correlating internal and external degrees of freedom. Its eigenstates take the form
\begin{equation}
\vert \Psi_{k} \rangle = \sum_{n} e^{\ii k z_{n}} \left[ e^{-\ii k_{c}z_{n}} c_{+}(k) \hat{\sigma}_{+g}^{(n)} + e^{\ii k_{c}z_{n}} c_{-}(k) \hat{\sigma}_{-g}^{(n)} \right]  \vert g \rangle^{\otimes N}
\end{equation}
as can be shown by direct substitution. Here, $k$ refers to the quasimomentum and is restricted to the Brillouin zone, $k \in [-\pi /a, \pi/a)$. These eigenvectors satisfy the eigenvalue equation
\begin{align} \label{eq_eigenstates}
\hat{I}(a,k_{c}) \vert \Psi_{k} \rangle = e^{i ka} \vert \Psi_{k} \rangle \, .
\end{align} 

Since the helical operator and the Hamilonian  $\tilde{\mathcal{H}}$ commute, they can be diagonalized simultaneously. We begin with the eigenstate $\vert \Psi_{k} \rangle $ that satisfies the eigenvalue equation
\begin{equation}
\tilde{\mathcal{H}} \vert \Psi_{k} \rangle = \hbar [\omega(k) - \tfrac{1}{2}\ii\Gamma(k)] \vert \Psi_{k} \rangle \, .
\end{equation}
and obtain the probability amplitudes $c_{\pm}(k)$. By projecting this equation into the $\vert e_{\pm}^{(n)} \rangle$ vectors, these amplitudes are shown to satisfy the matrix equation
\begin{equation}\label{eq:final_eigen}
[\omega(k) - \tfrac{1}{2}\ii\Gamma(k) - \tfrac{1}{2}(\Delta + \tfrac12\delta + \ii \Gamma_{0}){1} - \tfrac{1}{4} A  ] \left( \begin{array}{c}
c_{+}\\
c_{-}
\end{array} \right) = 0
\end{equation}
with 
\begin{equation}
A = \left( \begin{array}{cc}
 \delta \cos \theta - 4\tilde{\mathcal{K}}(k-k_{c}) &  \delta \sin \theta \\
 \delta \sin \theta & -\delta\cos \theta - 4\tilde{\mathcal{K}}(k+k_{c})
\end{array} \right) \, .
\end{equation}
In the last step  we used that the collective response of the atoms depends on their relative distance only. This allowed us to write the contribution for the $n$th atom as 
\begin{equation}
\sum_{m} \mathcal{K}_{\pm \pm}^{nm} e^{\ii (k \mp k_{c})(z_{m}-z_{n})} = \tfrac{\ii}{2} \Gamma_{0} + \tilde{\mathcal{K}}(k\mp k_{c}) \, ,
\end{equation}
which same value for each site in an infinite chain. The Fourier transform $\tilde{\mathcal{K}}(k)$ is given explicitly in the main text.  

The eigenvectors and eigenvalues of the $2\times 2$ matrix Eq.~(\ref{eq:final_eigen}) determine the probability amplitudes $c_{\pm}(k)$ and define the eigenstates of $\tilde{\mathcal{H}}$. They are written explicitly in Eqs.~(6) and~(7) of the main text.